\documentclass[aps,prb,preprintgroupedaddress,eqsecnum,draft,showpacs,intlimits,amsmath,amssymb]{revtex4}
\usepackage{bm}
\usepackage[final]{graphicx}
\usepackage{longtable}
\newcommand{\be}{\begin{equation}}
\newcommand{\ee}{\end{equation}}
\newcommand{\bea}{\begin{eqnarray}}
\newcommand{\eea}{\end{eqnarray}}

\begin{document}

\title{Surface critical behavior of semi-infinite systems with cubic anisotropy
at the ordinary transition.}

\author{Z. Usatenko$~{^{1,*}}$ and J. Spa\l{}ek$~{^{2}}$}

\affiliation{$^{1}$Institute for Condensed Matter Physics of the
National Academy of Sciences of Ukraine, 79011 Lviv, Ukraine}
\email{pylyp@ph.icmp.lviv.ua} \affiliation{$^{2}$ Marian
Smoluchowski Institute of Physics, Jagiellonian University, 30-059
Krakow, Poland}

\vspace{0.1cm}
\date{\today}

\begin{abstract}

The critical behavior at the ordinary transition in semi-infinite
$n$-component anisotropic cubic models is investigated by applying
the field theoretic approach in $d=3$ dimensions up to the
two-loop approximation. Numerical estimates of the resulting
two-loop series expansions for the critical exponents of the
ordinary transition are computed by means of Pad\'e resummation
techniques. For $n<n_{c}$ the system belongs to the universality
class of the isotropic $n$-component model, while for $n>n_{c}$
the cubic fixed point becomes stable, where $n_{c}<3$ is the
marginal spin dimensionality of the cubic model. The obtained
results indicate that the surface critical behavior of the
semi-infinite systems with cubic anisotropy is characterized by a
new set of surface critical exponents  for $n>n_{c}$.

\end{abstract}
\vspace{0.2cm} 

\maketitle

\renewcommand{\theequation}{\arabic{section}.\arabic{equation}}
\section{Introduction}
\setcounter{equation}{0}

The investigation of the critical behavior of real systems is an
important task of the condensed matter theory. The critical
behavior in such systems as polymers, easy-axis ferromagnets,
superconductors, as well as superfluid $^{4}{\mathrm He}$,
Heisenberg ferromagnets, and quark-gluon plasma is described by
isotropic $O(n)$ model with $n=0,1,2,3$ and $4$, respectively, and
has been analyzed in the framework of different theoretical and
numerical approaches.

Investigation of the critical behavior of real cubic crystals is
one of the topics of extensive theoretical work during the last
three decades. In crystals, due to their crystalline structure,
some kind of anisotropy is always present. One of the simplest
examples is the cubic anisotropy. A typical model of the critical
behavior of such systems is the model with a cubic term
$\frac{v_{0}}{4 !}\sum_{i=1}^{n}\phi_{i}^{4}$ added to the usual
$O(n)$ symmetric $\frac{u_{0}}{4 !}
(\sum_{i=1}^{n}\mid\phi_{i}\mid^{2})^2$ term
\cite{Ahar73,KW73,W73}. This $n$-component cubic model is a
particular case of an $mn$-component model \cite{BGZ74}, with the
cubic anisotropy at $m=1$. The model exhibits several types of
continuous and first-order phase transitions depending on the
number of spin components $n$, space dimensionality $d$, and the
sign of the cubic coupling constant $v_{0}$. The cubic models are
widely applied to the study of magnetic and structural phase
transitions. In the limiting case of $n\to 0$,
 it describes the critical behavior of random Ising-like systems \cite{GL76}. The
 case $m=0$ and $n=0$ formally describes the critical behavior of long
 flexible polymer chains in good solvents as model of self-avoiding walks (SAW)
 on a regular lattice, with short range
 correlated quenched disorder. As has been shown by Harris \cite{Harris83} and Kim \cite{Kim}, the short range correlated (or random uncorrelated pointlike) disorder is
irrelevant for such model.
 The case $m=1$ and $n\to \infty$ corresponds to the Ising model with
 equilibrium magnetic impurities \cite{Aharony73}.

 Depending on the sign of the cubic coupling constant $v_{0}$, the two types of order are
possible: along the diagonals the type [1,1,...,1] of a hypercube
in $n$ dimensions for $v_{0}>0$ or along the easy axes of the type
[1,0,...0] for $v_{0}<0$. In the latter case the system can
undergo the first order phase transition, as was confirmed in
experiments \cite{Sznajd84}. In the present work we are concerned
with the case $v_{0}>0$.

The presence of a surface leads to the appearance of additional
complications. The source of these problems is connected with both
the loss of translational invariance and the presence of a
boundaries. General reviews on surface critical phenomena are
given in Refs. \cite{B83,D86,D97}. The simplest model of critical
phenomena in systems with a single planar surface is the
semi-infinite model \cite{B83}. As it is known
\cite{BH72,B83,D86}, the phase diagram of such model is richer
than that of its bulk correspondant. In general case of pure
semi-infinite model with continuous $O(n)$ symmetries, there are a
surface- and bulk-disordered phases (SD and BD, respectively), as
well as either the surface-ordered, bulk-disordered phase (SO and
BD, respectively) and a surface-ordered, bulk-ordered phase (SO
and BO). The surface phase can actually occur, if $d>2$ and $n=1$
or $d>3$ and $n>2$. The boundaries between the phases are the
lines of surface, ordinary, and extraordinary transitions which
meet at a multicritical point
$(m_{0}^{2},c_{0})=(m_{0c}^{2},c_{sp}^{*})$, representing the
special transition and called the special point. Each of the
above mentioned transitions is characterized by own fixed point.
The constant $c_{0}$ is related to the surface enhancement, which
measures the enhancement of the interactions at the surface. The
coupling $m_{0}$ is defined in Eq.(\ref{1}) below. The
nonexistence of this ordered surface phase assume the occurrence
of extraordinary transitions. In the case $n=2$ and $d=3$ a
surface transitions of the Kosterlitz-Thouless type is present. We
do not consider this type of transition and extraordinary ones,
because they have a different nature than the special and ordinary
transitions.

In general, there are different surface universality classes,
defining the critical behavior in the vicinity of the system
boundaries, at temperatures close to the bulk critical point
($\tau=(T-T_{c})/T_{c}\to 0$). Each bulk universality class
divides into several distinct surface universality classes. Three
surface universality classes, called respectively ordinary
($c_{0}\to \infty$), special ($c_{0}=c_{sp}^{*}$) and
extraordinary ($c_{0}\to -\infty$), are known \cite{D86,D97,DD81}.

In order to investigate the critical behavior of real cubic
crystals we must take into account that two types of anisotropy
can be present for such systems. The first one is a bulk
anisotropy, which can be included into the consideration with a
help of above mentioned cubic term. The other one is a surface
anisotropy which arise as a consequence of presence the bulk cubic
anisotropy (see Appendix A for details). In the present paper we
are interested in investigation of the critical behavior only at
the ordinary transition, where the surface orders simultaneously
with the bulk. In this case, as was founded by Diehl and
Eisenriegler \cite{DE82,DE84}, the surface anisotropy is
irrelevant.

Theory of critical behavior of individual surface universality
classes is very well developed for pure isotropic systems
\cite{D97,DD81,DD83,DDE83,DC91,DSh98}, the systems with quenched
surface-enhancement disorder \cite{DN90,PS98,D98} and the systems
with a random quenched bulk disorder for both the ordinary and the
special surface transitions \cite{UShH00,UH02}. General
irrelevance-relevance criteria of the Harris type for the systems
with quenched short-range correlated surface-bond disorder were
predicted in \cite{DN90} and confirmed by Monte-Carlo calculations
\cite{PS98,ILSS98}. Moreover, it was established that the surface
critical behavior of semi-infinite systems with quenched bulk
disorder is characterized by the new set of surface critical
exponents in comparison with the case of pure systems
\cite{UShH00,UH02}.

The remainder of this paper is organized as follows. Section II
contains the description of model and further useful background.
In Section III the renormalization group approach is described.
Section IV contains the calculations of the surface
renormalization factor $Z_{\partial\varphi}$ and surface critical
exponent $\eta_{\partial\varphi}$ by applying the field theoretic
approach directly in $d=3$ dimensions, up to the two-loop order.
The numerical estimates of the resulting two-loop series
expansions for the critical exponents of the ordinary transition
are presented in Section V. The calculations are performed by
means of the Pad\'e resummation techniques for the cases $n=3,4,8$
and for the case of $n\to\infty$, which corresponds to the Ising
model with equilibrium magnetic impurities. Section VI contains
concluding remarks. The Appendix A contains the
Landau-Ginzburg-Wilson functional for a semi-infinite systems of
spins with cubic anisotropy. The Appendix B contains the standard
surface scaling relations for the case $d=3$.

\renewcommand{\theequation}{\arabic{section}.\arabic{equation}}
\section{The model}
\setcounter{equation}{0}

The effective Landau-Ginzburg-Wilson Hamiltonian of the $n$-vector
model with cubic anisotropy in the semi-infinite space is given by
(see Appendix A)
\begin{eqnarray}
H(\vec{\phi}) & = &\int_{0}^\infty dz \int d^{d-1}r [\frac{1}{2}
\mid \nabla\vec{\phi} \mid ^{2} +
\frac{1}{2} m_{0}^{2}\mid \vec{\phi} \mid^{2}\nonumber\\
&+& \frac{1}{4!} v_{0} \sum_{i=1}^{n} |\phi_{i}|^{4}+
\frac{1}{4!}u_{0} (\sum_{i=1}^{n} \mid
\phi_{i}\mid^{2})^{2}]\label{1} \end{eqnarray} where
$\vec{\phi}(x)=\{\phi_{i}(x)\}$ is an $n$-vector field with the
components $\phi_{i}(x)$, $i=1,...,n$. Here $m_{0}^2$ is the "bare
mass", representing a linear measure of the temperature difference
from the critical point value. The parameters $u_{0}$ and $v_{0}$
are the usual "bare" coupling constants $u_{0} > 0$ and $v_{0} >
0$. It should be mentioned that the $d$-dimensional spatial
integration is extended over a half-space $I\!\!R^d_+\equiv\{{\bf
x}{=}({\bf r},z)\in I\!\!R^d$, with ${\bf r}\in I\!\!R^{d-1}$ and
$z\ge 0\}$, bounded by a planar free surface at $z=0$. The fields
$\phi_{i}({\bf r},z)$ satisfy the Dirichlet boundary condition
$\phi_{i}({\bf r},z)=0$ at $z=0$ in the case of ordinary
transition, and the Neumann boundary condition
$\partial_{n}\phi_{i}({\bf r}, z)=0$ at $z=0$ in the case of
special transition \cite{DD81,DDE83}. The model defined in
(\ref{1}) is translationally invariant in directions parallel to
the external surface, $z=0$. Thus, we shall use mixed
representation, i.e. Fourier representation in $d-1$ dimensions
and real-space representation in $z$ direction. Therefore, as was
mentioned above \cite{DE82,DE84} and shown in Appendix A, no
specific surface term will appear in the case of ordinary
transition, when the Dirichlet boundary conditions on the surface
are assumed.

The added cubic term breaks the $O(n)$ invariance of the model,
leaving a discrete cubic symmetry. The model (\ref{1}) has four
fixed points: the trivial Gaussian, the Ising one in which the $n$
components are decoupled, the isotropic ($O(n)$-symmetric), and
the cubic fixed points. The Gaussian and Ising fixed points are
never stable for any number of components $n$. For isotropic
systems, the $O(n)$-symmetric fixed point is stable for $n<n_{c}$,
whereas for $n>n_{c}$ it becomes unstable. Here $n_{c}$ is the
marginal spin dimensionality of the cubic model, at which the
isotropic and cubic fixed points change stability, i.e. for
$n>n_{c}$, the cubic fixed point becomes stable. The
$O(n)$-symmetric fixed point is tricritical. At $n=n_{c}$, the two
fixed points should coincide, and logarithmic corrections to the
$O(n)$-symmetric critical exponents are present. The calculation
of the critical marginal spin dimensionality $n_{c}$ is the
crucial point in studing the critical behavior in
three-dimensional cubic crystals. Different results for $n_{c}$
have been published in a series of works in which different
methods have been used.
 In the framework of the
field-theoretical RG analysis the one-loop and three-loop
approximations at $\epsilon=1$ lead to the conclusion that $n_{c}$
should lie between 3 and 4 \cite{KetleyW73,AharonyBruce74}, and
the cubic ferromagnets are described by the Heisenberg model. On
the other hand, by using the field theoretic approach directly in
$d=3$ dimensions up to the three-loop approximation, it has been
found that $n_{c}=2.9$ \cite{MayerSokolov87,Sh89}. Similar
conclusions were obtained in \cite{YalabicHoughton77}, where it
was found that $n_{c}=2.3$. The calculations performed by Newman
and Riedel \cite{NR82} with the help of the scaling-field method,
developed by Goldner and Ridel for Wilson's exact momentum-space
RG equations, have given for $d=3$ the value $n_{c}=3.4$.
Field-theoretical analysis, based on the four-loop series in three
dimensions \cite{MSSh89,V00} and results of the five-loop
$\epsilon=4-d$ expansion \cite{V00,KSch95,ShAS97} suggest that
$n_{c}\le 3$. Recently, a very precise six-loop result for the
marginal spin dimensionality of the cubic model, $n_{c}=2.89(4)$,
was obtained in the framework of the 3D field-theoretic approach
\cite{CPV00}. Thus, it was finally established that the critical
behavior of the cubic ferromagnets is not described by the
isotropic Heisenberg Hamiltonian, but by the cubic model, at the
cubic fixed point. However, it was found that the difference
between the values of the bulk critical exponents at the cubic and
the isotropic fixed points is very small, i.e. it is hard to
determine this difference experimentally. Nevertheless, the
recently obtained results stimulated us to perform the analysis of
the {\it surface critical behavior} of semi-infinite n-component
anisotropic cubic model, and to determine corresponding surface
critical exponents.

\section{Renormalization}

The fundamental two-point correlation function of the free static
theory corresponding to (\ref{1}) is defined by the Dirichlet
propagator:
\be \langle \varphi_i(r,z)\varphi_j(0,z')\rangle_0 =
G_{D}(r;z,z') \delta_{ij}.
 \ee
In the mixed $pz$ representation the Dirichlet propagator is
\be\label{prop} G_{D}(p;z,z')={1
\over2\kappa_0}\left[e^{-\kappa_0|z-z^\prime|}-
e^{-\kappa_0(z+z^\prime)}\right], \ee where the standard
notationis used and, $\kappa_0=\sqrt{p^2+m_0^2}$. The propagator
vanishes identically when at least one of its $z$ coordinates is
zero, because we have assumed the Dirichlet boundary conditions.
Consequently, all the correlation functions involving at least one
field at the surface vanish. This property holds for both the free
and the renormalized theories \cite{D86}.

In fact the critical surface singularities at the ordinary
transition can be extracted by studying the nontrivial in this
case correlation function involving the ({\it inner}) normal
derivatives of the fields at the boundary, $\partial_n\phi(r)$
\cite{DD80,DD81,DDE83}. Actually, in order to obtain the
characteristic exponent $\eta_\parallel^{ord}$ of surface
correlations, it is sufficient to consider a correlation function
with two normal derivatives of boundary fields, i.e.
\begin{equation} {\cal G}_2(p)=\left\langle {\partial\over\partial
z}\left.\varphi(p, z)\right|_{z=0} {\partial\over\partial
z'}\left.\varphi(-p, z')\right|_{z'=0} \right\rangle,
\end{equation}
where the fields $\varphi(p,z)$ are the Fourier transforms of the
fields $\varphi(r,z)$ in the $(d-1)$ dimensional parallel to
surface space. ${\cal G}_2(p)$ is a parallel Fourier transform of
the corresponding two-point function ${\cal G}_2(r)$ in direct
space. At the critical point ${\cal G}_2(p)$ behaves as
$p^{-1+\eta_\parallel^{ord}}$. It reproduces the leading critical
behavior of a two-point function $G_2(p)=\langle\varphi(p,
z)\varphi(-p, z^{'})\rangle$ in the vicinity of the boundary
plane. The surface critical exponent $\eta_\parallel^{ord}$ is
provided by the scaling dimension of the boundary operator
$\partial_n\varphi(r)$.

The surface correlation function exponent $\eta_\parallel^{ord}$
in the semi-infinite systems with cubic anisotropy differs from
its corresponding value for the isotropic semi-infinite system.
The remaining surface critical exponents of the ordinary
transition can be determined through the surface scaling laws
\cite{D86}. (see Appendix B).

In the present formulation of the problem, the renormalization
process for the cubic anisotropic system is essentially the same
as in the isotropic case \cite{D86,DSh98}. Explicitly, the
renormalized bulk field and its normal derivative at the surface
should be reparametrized by different uv-finite renormalization
factors $Z_{\varphi}(u,v)$ and $Z_{\partial{\varphi}}(u,v)$
\begin{equation}
\!\varphi_{R}(x){=}Z_\varphi^{-{1\over 2}}\,\varphi(x)
\quad\mbox{and}\quad
\big(\partial_n\varphi(r)\big)_R{=}Z_{\partial{\varphi}}^{-{1\over
2}}\partial_n\varphi(r),\nonumber \end{equation} and renormalized
correlation functions involving $N$ bulk fields and $M$ normal
derivatives are
\begin{equation}
{\cal G}_R^{(N,M)}(p;m,u,v)
=Z_\varphi ^{-{N\over 2}}Z_{\partial{\varphi}} ^{-{M\over 2}} {\cal
G}^{(N,M)}(p;m_{0},u_0,v_0)
\end{equation}
for $(N,M)\ne (0,2)$. In  order to remove the ultraviolet (uv)
singularities of the correlation function ${\cal G}^{(0,2)}$ with
two surface operators $(N,M)=(0,2)$ in the vicinity of the
surface, an additional, {\it additive} renormalization
(zero-momentum subtraction) is required, so that
\begin{equation}\label{addren} {\cal
G}_R^{(0,2)}(p)=Z_{\partial{\varphi}}^{-1} \left[{\cal
G}^{(0,2)}(p)- {\cal G}^{(0,2)}(p{=}0)\right]\,.
\end{equation}

The typical bulk uv singularities, which are present in the
correlation function  ${\cal G}^{(0,2)}$, are subtracted via the
standard mass renormalization of the massive infinite-volume
theory. It also relates to coupling constants, for which standard
vertex renormalization of coupling constants takes place.

The surface renormalization factor $Z_{\partial{\varphi}}(u,v)$ can be
conveniently obtained from the consideration of the boundary two-point
function ${\cal G}^{(0,2)}$,
\begin{equation}\label{zn}
Z_{\partial{\varphi}}=-\lim_{p\to 0}{m\over p}{\partial\over\partial
p}\,{\cal G}^{(0,2)}(p)\,. \end{equation}
A standard RG argument involving an inhomogeneous Callan-Symanzik equation
yields the anomalous dimension of the operator $\partial_n\varphi(r)$
\begin{eqnarray}\label{etan}
\eta_{\partial{\varphi}}&=&m\!{\partial\over \partial m}
\left.\ln\!{Z_{\partial{\varphi}}}\,\right|_{F\!P}\\
      &=&\beta_u(u,v){\partial\ln Z_{\partial{\varphi}}(u,v)\over \partial
      u}+
 \left.\beta_v(u,v){\partial\ln Z_{\partial{\varphi}}(u,v)\over \partial
 v}\right |_{F\!P}\,.\nonumber
\end{eqnarray}
"FP" indicates here that the above value should be calculated at the
infrared-stable cubic fixed point of the underlying bulk theory,
$(u,v)=(u^*,v^*)$. The surface critical exponent $\eta_{\parallel}^{ord}$ at
the ordinary transition is then given by
\begin{equation}
\eta_{\parallel}^{ord}=2+\eta_{\partial{\varphi}}\,.
\end{equation}

\renewcommand{\theequation}{\arabic{section}.\arabic{equation}}
\section{Perturbation theory up to two-loop approximation}
\setcounter{equation}{0}

After performing the mass and additive renormalization of the correlation
function ${\cal G}^{(0,2)}(p)$
and carrying out the integration of Feynman integrals by analogy
with Ref.\cite{DSh98,UShH00}, we obtain for renormalization factor

\begin{equation}
Z_{\partial\varphi}(\bar u_0,\bar v_0)=1+{\bar t_1^{(0)}\over 4} +\bar
t_2^{(0)} C,
\end{equation}

 where the constant $C$ follows from the two-loop (melon-like diagrams)
contribution to the correlation function and has the value,
\begin{equation}\label{co}
C\simeq{107\over 162}-  {7\over 3} \ln{4\over 3}-0.094299\simeq
-0.105063.
\end{equation}
 The coefficients ${\bar t_{1}^{(0)}}$ and ${\bar t_{2}^{(0)}}$ are the weighting factors
 belonging to one- and two-loop (melon-like) diagrams in
 the Feynman diagrammatic expansion of the correlation function ${\cal
 G}^{(0,2)}(p)$, and equal

\begin{mathletters}\label{ti}
\begin{eqnarray}
&-&\frac{{\bar t}_1^{(0)}}{2},\quad\mbox{with}\quad
{\bar t}_1^{(0)}=\frac{n+2}{3}\,{\bar u_0}+ {\bar v_{0}}\,,\\
&&\frac{{\bar t}_2^{(0)}}{6},\quad\mbox{with}\quad {\bar
t}_2^{(0)}=\frac{n+2}{3}\, {\bar u_{0}}^{2}+ {\bar v_{0}}^{2}+
2{\bar v_{0}}{\bar u_{0}}\,. \end{eqnarray} \end{mathletters}

The factors ${\bar t}_{1}^{(0)}$ and ${\bar t}_{2}^{(0)}$
 follow from the standard symmetry properties of Hamiltonian (\ref{1}).
Here the renormalization factor $Z_{\partial \varphi}$ is
expressed as a second-order series expansion in powers of {\it
bare} dimensionless parameters $\bar u_0=u_0/(8\pi m)$ and $\bar
v_0=v_0/(8\pi m)$. As it is usual in {\it super}renormalizable
theories, the renormalization factor expressed in terms of
unrenormalized coupling constants, is finite.

As a next step, the vertex renormalizations should be carried out.
To the present accuracy, they are
\begin{mathletters}
\begin{eqnarray}
\bar{u}_{0}&=&\bar{u}\Big(1+\frac{n+8}{6}\,\bar{u}+\bar{v}\Big),\\
\bar{v}_{0}&=&\bar{v}\Big(1+\frac{3}{2}\,\bar{v}+2\, \bar{u}\Big)\,.\label{33}
\end{eqnarray}
\end{mathletters}
As known, the vertex renormalization at $d=3$ is a finite
reparametrization. All relevant singularities have been removed
already after the mass renormalization and taking into account the
special bubble-graph combinations emerging in the theory with {\it
Dirichlet} propagators. Thus we obtain a modified series expansion
up to two-loop approximation
\begin{eqnarray}
Z_{\partial \varphi}(\bar u, \bar v)&=&
1+{n{+}2\over 12}\,\bar u +{\bar v\over 4}
+{n{+}2\over 3}\left(C+{n{+}8\over 24}\right)\,\bar u^2\nonumber\\
&+&\left(C+{3\over 8}\right)\,\bar v^2+2\left(C+{n{+}8\over 24}\right)\,\bar u\bar v\,.
\end{eqnarray}

Combining the renormalization factor $Z_{\partial \varphi}(\bar u,
\bar v)$ together with the one-loop pieces of the beta functions :
$\beta_{\bar u}(\bar u, \bar v)=-\bar u\Big(1-{n{+}8\over 6}\;\bar
u-\bar v\Big)$ and  $\beta_{\bar v}(\bar u, \bar v)=-\bar
v\Big(1-{3\over 2}\,\bar v-2\bar u\Big)$ and inserting them into
Eq.(\ref{etan}), we obtain the desired series expansion for
$\eta_{\partial \varphi}$,

\begin{eqnarray}\label{etafin}
&&\eta_{\parallel}(u,v)=2-{n{+}2\over {2 (n{+}8)}}\,u-{v\over 6}\\
&&-24{(n{+}2)\over (n{+}8)^2}{\cal C}(n) u^2-{8\over 9}{\cal
C}(1) v^2 -{16\over n{+}8}{\cal C}(n) uv ,\nonumber
\end{eqnarray}
where ${\cal C}(n)$ is a function of the order-parameter
components number $n$, and is defined as
\begin{equation}
{\cal C}(n)=C+{n{+}14\over 96},
\end{equation}
whereas the renormalized coupling constants $u$ and $v$, normalzed
in a standard fashion are $u{=}{n+8\over 6}{\bar{u}}$ and
$v{=}{3\over 2}{\bar{v}}$.

Eq.(\ref{etafin}) supplies our result for the critical exponent of
surface correlation function for the model with the effective
Hamiltonian of the Landau-Ginzburg-Wilson type with cubic
anisotropy in the semi-infinite space (\ref{1}) with general
number $n$ of order parameter components.

The knowledge of $\eta_{\parallel}$ gives possibility to calculate
the other surface critical exponents through the scaling
relations. For convenience, from now on we omit the superscript
{\it ord} for the surface critical exponents.

The critical exponents should be calculated for different $n$
($n=3,4,8$, and $n \to\infty$) at the standard infrared-stable
cubic fixed (FP) points of the underlying bulk theory, as is
usually accepted in the massive field theory. As was mentioned
above, in the cases $n<n_{c}$ the cubic ferromagnets are described
by the Heisenberg isotropic Hamiltonian at the $O(n)$-symmetric
fixed point.

In the case of the replica limit $n\to 0$ we obtain from
(\ref{etafin}) the series expansion of $\eta_{\parallel}^{r}$ for
{\it{semi-infinite random Ising-like}} systems. This case was
investigated in detail of one of us previously \cite{UShH00}.

\section{Numerical results}

In order to obtain the full set of surface critical exponents for
the ordinary transition in systems with cubic anisotropy, we
substitute the expansion (\ref{etafin}) for $\eta_{\parallel}$
into the standart scaling-law expressions for the surface
exponents (see Appendix 1).

For each of the above mentioned surface critical exponents of the
ordinary transition  we obtain for $d=3$ a double series expansion
in powers of $u$ and $v$, truncated at the second order. As it is
known \cite{mit3,B87,Mc94,AMR00}, power series expansions of this
kind are generally divergent due to a nearly factorial growth of
expansion coefficients at large orders of perturbation theory. In
order to perform the analysis of these perturbative series
expansions and obtain accurate estimates of the surface critical
exponents, a powerful resummation procedure must be used. One of
the simplest ways is to perform the double Pad\'e-analysis
\cite{B75}. This should work well when the series behaves in
lowest orders "in a convergent fashion".

The results of our calculations of the surface critical exponents
of the ordinary transition for various values of $n=3,4,8,\infty$
at the corresponding cubic fixed points are presented in Tables
1-5. Unfortunately, the second-order (p=2) analysis of
perturbative series \cite{mitSh88} gives the cubic fixed point
with coordinates $u_{0}=1.5347$ and $v_{0}=-0.0674$ at $n=3$ for
the 3D model.
  The analysis of the eigenvalues of the stability matrix shows
that in the frames of the two-loop approximation the cubic fixed
point at $n=3$ is unstable and the $O(n)$-symetric fixed point is
stable. But, the estimates of
 the marginal spin dimensionality of the cubic model $n_{c}$ in the
 frames of three-loop \cite{MayerSokolov87,Sh89},
 four-loop \cite{MSSh89,V00}, five-loop $\epsilon=4-d$ expansion
\cite{KSch95,ShAS97,V00} and six-loop study \cite{CPV00} show that
the cubic ferromagnets are not described by the Heisenberg
isotropic model, but by the cubic model at the stable cubic fixed
point. Higher precision six-loop field-theoretical analysis
\cite{CPV00} give the value of the marginal spin dimensionality of
the cubic model equal to $n_{c}=2.89(4)$.  In accordance with this
we use the cubic fixed point of the higher $p=3$ order of
perturbative series for obtaining the set of surface critical
exponents at $n=3$. For estimation of the reliability of the
obtained results we performed calculations at the cubic fixed
point of the $p=6$ order in Table 2. We obtained that difference
in these two cases are approximatively: $0.5 \%$ for
$\eta_{\parallel}$, $0.4 \%$ for $\eta_{\perp}$, $0.8 \%$ for
$\Delta_{1}$, $0.1 \%$ for $\beta_{1}$, $6 \%$ for $\gamma_{11}$,
$0.2 \%$ for $\gamma_{1}$, $0.2 \%$ for $\delta_{1}$, and $0.8 \%$
for $\delta_{11}$. The obtained results indicate that the
difference in the ways of the $\beta$ functions resummation have
no essential influence on the values of the surface critical
exponents and that the results obtained in the frames of the
two-loop approximation are stable and reliable. The surface
critical exponents of the ordinary transition for $n=4,8$ and
$n\to\infty$ were calculated at the standard infrared-stable cubic
fixed (FP) points of the underlying bulk theory, as it is usually
accepted in the massive field theory.

 The quantities $O_{1}/O_{2}$ and $O_{1i}/O_{2i}$ represent the ratios of
magnitudes of first-order and second-order perturbative corrections
appearing in direct and inverse series expansions. The larger (absolute)
value of these ratios indicate the better apparent convergence of
truncated series.

The values $[p/q]$ (where $p,q=0,1$) are simply Pad\'e
approximants which represent the partial sums of the direct and
inverse series expansions up to the first and the second order.
The nearly diagonal two-variable rational approximants of the
types $[11/1]$ and $[1/11]$ give at $u=0$ or $v=0$ the usual
$[1/1]$ Pad\'e approximant \cite{B75}. As it is easy to see from
Tables 1-5, the values of $[11/1]$ and $[1/11]$ Pad\'e aproximants
do not differ significantly between themselves. We consider these
values as the best we could achieve from the available knowledge
about the series expansions in the two-loop approximation scheme.
Thus, our final results are presented in the last columns of the
Tables 1-5. Their deviations from the other second-order estimates
might serve as a rough measure of the achieved numerical accuracy.
 As it is easy to see, the obtained results indicate
about good stability of the results calculated
in the frames of the two-loop approximation scheme.

The results for surface critical exponents of
semi-infinite model with cubic anisotropy, calculated at the cubic fixed
point are different from the results for surface critical exponents of
standart semi-infinite $n$-component model (see
\cite{DD81,DD83,DSh94,DSh98}).

If $n<n_{c}$, the cubic fixed point is unstable and
the cubic term in the Hamiltonian (\ref{1}) becomes irrelevant. In this case
the isotropic fixed point is stable and the system is described by the
simple $O(n)$-symmetric model in 3D. The corresponding
 surface critical exponents can be
calculated from the series, presented in \cite{DSh98}.

As was indicated previously, in the limit $n\to 0$, the cubic model
(\ref{1}) with $u_{0}<0$ and $v_{0}>0$ describes the semi-infinite
Ising-like systems with random bulk disorder. The investigation of the
ordinary transition for such kind of systems was presented in
\cite{UShH00}.

\section{Concluding remarks}

We have studied {\it ordinary transition} for a semi-infinite
systems with cubic anisotropy by applying the field theoretic
approach directly in $d=3$ dimensions, up to the two-loop
approximation. We have performed a double Pad\'e analysis of the
resulting perturbation series for the surface critical exponents
of the ordinary transition for various $n=3,4,8,\infty$, in order
to find the best numerical estimates. We find that at $n>n_{c}$
the surface critical exponents of the ordinary transition in
semi-infinite systems with cubic anisotropy belong to the cubic
universality class.

In order to obtain more precise numerical estimates for the case
of 3D dimensional cubic crystal with $n=3$, a further theoretical
investigation of the asymptotic surface critical behavior of
semi-infinite cubic systems would be highly desirable within the
framework of higher-order RG approximations.

We suggest that the obtained results could stimulate further experimental
and numerical investigations of the surface critical behavior of random
systems and systems with cubic anisotropy.

\section*{Acknowledgments}

 One of us (Z.U.) was supported in part by the Queen Jadwiga Fellowship
 of the
 Jagiellonian University, Krakow, Poland. The second author (J.S.)acknowledges
 the support of the State Committee for Scientific Research (KBN), Grant
 N 2P03B05023, as well as the Fellowship of the Foundation for Science (FNP).   .

\renewcommand{\theequation}{A1.\arabic{equation}}
\section*{Appendix A. The Landau-Ginzburg-Wilson functional for the
semi-infinite systems with cubic anisotropy}
\setcounter{equation}{0}

We outline the main steps of deriving the Landau-Ginzburg-Wilson
functional for the case of Heisenberg  ferromagnet and discuss
briefly the surface term leading to the surface boundary
conditions. We perform the calculation for a simple cubic
structure and start from the mean-field approximation for the
Heisenberg exchange interaction, as the cubic-anisotropy term can
be added afterwards. Additionally, we limit ourselves to
one-component theory, as the generalization to the $n$-component
version is not important to the essence of the principal argument.

Suppose we have the system of localized spins of magnitude $S$ and
described by the Heisenberg Hamiltonian, which in the applied
field $h$ takes the form

\begin{equation}
H=-\frac{1}{2}\sum_{i\ne j}
J_{ij}{\bf{S}}_{i}\bullet{\bf{S}}_{j}-h\sum_{i}
S_{i}^{z}.\label{a1}
\end{equation}
In the mean-field approximation
\begin{equation}
{\bf{S}}_{i}\bullet{\bf{S}}_{j}=<{\bf{S}}_{i}>\bullet
{\bf{S}}_{j}+<{\bf{S}}_{j}>\bullet
{\bf{S}}_{i}-<{\bf{S}}_{i}>\bullet<{\bf{S}}_{j}>,\label{a2}
\end{equation}
and for the spin quantization axis taken as the z-axis. We can
write the free energy in the form
\begin{equation}
F=-{\tilde{N}}k_{B}T \ln \frac{\sinh(\beta
h_{i}(S+1/2))}{\sinh(\beta h_{i}/2)}+\frac{1}{2}\sum_{i\ne
j}J_{ij}<S_{i}^{z}><S_{j}^{z}>,\label{a3}
\end{equation}
where $h_{i}=\sum_{j}J_{ij}<S_{j}^{z}>+h$ is the effective field
acting on $S_{i}^{z}$, $\beta=(k_{B}T)^{-1}$ is the inverse
temperature in energy units, and $\bar{N}$ is the total number of
spins.

The constant term can be rewritten in the first nontrivial order
the continuum-medium approximation as
\begin{equation}
\frac{1}{2}\sum_{j}J_{ij}<S_{i}^{z}><S_{j}^{z}>\simeq\left.\frac{1}{2}
J_{0}<S^{z}({\bf{x}})>\right|_{{\bf{x}}={\bf{R}}_{i}}
+\left.\frac{1}{2}a_{0}^{2}<S^{z}({\bf{x}})>\nabla^{2}<S^{z}({\bf{x}})>\right|_{{\bf{x}}={\bf{R}}_{i}},\label{a4}
\end{equation}
where $J_{0}=\sum_{j}J_{ij}$, $a_{0}$ is the lattice constant, and
$i={\bf{R}}_{i}$ denotes here the lattice site position.

The expansion of the ratio of hyperbolic function in $y=\beta
h_{i}$ can be represented as
\begin{eqnarray}
\frac{\sinh(y(S+1/2))}{\sinh(y/2)}&\simeq&(2S+1)[1+\frac{1}{6}S(S+1)y^2\nonumber\\
&+&\frac{1}{360}S(3S^3+6S^2+2S-1)y^4]+O(y^6).\label{a5}
\end{eqnarray}
Hence, the free energy (per site) to the same approximation after
taking the continuum-medium limit reads
\begin{eqnarray}
\frac{F}{{\tilde{N}}}&=&\frac{F_{0}}{\tilde{N}}+\int d^{d}x
(\frac{3}{2S(S+1)}k_{B}(T-T_{c})\phi({\bf{x}})^2\nonumber\\
&+&\frac{1}{2}J_{0}a_{0}^2|\nabla
\phi({\bf{x}})|^2-\frac{1}{3}S(S+1)\beta h J_{0}\phi({\bf{x}})\nonumber\\
&+&\frac{\beta^3}{72}(S^2(S+1)^2-\frac{S}{5}(3S^3+6S^2+2S-1))(J_{0}\phi({\bf{x}})^4))\nonumber\\
&+&o(|\phi|^6)+o(h^2)+o(|\nabla \phi|^4), \label{a6}
\end{eqnarray}
with $\frac{F_{0}}{{\tilde{N}}}=-k_{B}T \ln(2S+1)$.
 The continuous
field $\phi({\bf{x}})$ expresses the limiting value of
$<S_{i}^{z}>$ per volume $a_{0}^{d}$.

Defining the mean-field critical temperature
$T_{c}=\frac{1}{3k_{B}}J_{0}S(S+1)$ and assuming that we can put
$T\simeq T_{c}$ in the last two terms we obtain the desired free
energy functional in the form
\begin{eqnarray}
\frac{F}{\tilde{N}}&=&\frac{F_{0}}{\tilde{N}}+\int d^{d}x
(\frac{A_{0}}{2}|\nabla\phi({\bf{x}})|^2+\frac{A_{1}}{2}(T-T_{c})|\phi({\bf{x}})|^2\nonumber\\
&+&\frac{A_{2}}{4!}|\phi({\bf{x}})|^4-h\phi({\bf{x}})),\label{a7}
\end{eqnarray}
where the constants are related in an obvious fashion to the
coefficients $m^{2}_{0}$ and $u_{0}$, when we divide $F$ by the
exchange stiffness constant $A_{0}$, which contains non-divergent
constants at the critical point.

Now, the boundary conditions appear when we derive the
Landau-Ginzburg equation in the form of Euler equation for
$\phi({\bf{x}})$. However, in such a situation specific surface
terms appear. The are two types of terms. First, is the gradient
term $A^{'}_{0}|\nabla\phi({\bf{r}},z=0)|^{2}$, since in
$J_{0}=\sum_{j}J_{ij}$ the spins above the surface are missing.
Second, the geometrical surface term of the form
$\frac{1}{2}c_{0}\int d^{d-1}r\phi^{2}({\bf{r}},z=0)$ may appear
where the $c_{0}$ is the surface enhancement constant. This is
becouse on the surface the role of the bulk cubic anisotropy is
taken over by the surface anisotropy, which in the first
nontrivial order has the form
$\frac{1}{2}\sum_{i=1}^{n}c^{i}_{0}\phi^{2}_{i}({\bf{r}},z=0)$. In
effect, the condition on the surface coming from the Euler
variational scheme takes the following form

\begin{equation}
\sum_{i=1}^{n}\int d^{d-1}r\{A^{'}_{0}{\vec{n}}\bullet\nabla
\phi_{i}({\bf{r}},z)+c^{i}_{0}\phi_{i}({\bf{r}},z)\} \delta
\phi_{i}({\bf{r}},z)|_{z=0}=0,\label{a8}
\end{equation}
where $\delta \phi_{i}({\bf{r}},z=0)$ is the variation of
$\phi_{i}({\bf{r}},z)$ on the surface and ${\vec{n}}$ is the
vector perpendicular to the surface. Therefore, we can choose the
boundary conditions in either way, namely
\begin{eqnarray}
1)&&\quad\quad \delta \phi_{i}({\bf{r}},z)|_{z=0}=0, \quad
i.e.\quad
\phi_{i}({\bf{r}},z)|_{z=0}=const=S_{0},\\
 2)&&\quad\quad
(\partial_{n}{\phi_{i}({\bf{r}},z)}+\frac{c^{i}_{0}}{A^{'}_{0}}\phi_{i}({\bf{r}},z))|_{z=0}=0.\label{a10}
\end{eqnarray}
In this paper we have selected the b.c. $1)$ with $S_{0}=0$, which
correspond to the Dirichlet boundary condition. In such a
situation, no specific surface term appears in the starting
functional (\ref{1}). However, it must be said that it is the
concrete experimental situation that determines the type of
boundary conditions to be taken into the theoretical analysis.

\renewcommand{\theequation}{A1.\arabic{equation}}
\section*{Appendix B. Scaling relations between the surface critical
exponents}
\setcounter{equation}{0}

The individual RG series expansions for other critical exponents
can be derived through standard surface scaling relations
\cite{D86} with $d=3$:
\begin{eqnarray}
&& \eta_{\perp} = \frac{\eta +
\eta_{\parallel}}{2}, \nonumber\\
&& \beta_{1} = \frac{\nu}{2}
(d-2+\eta_{\parallel}), \nonumber\\
&& \gamma_{11}=\nu(1-\eta_{\parallel}), \nonumber\\
&& \gamma_{1}= \nu(2-\eta_{\perp}), \label{sc}\\
&& \Delta_{1}= \frac{\nu}{2} (d-\eta_{\parallel}), \nonumber\\
&& \delta_{1} = \frac{\Delta}{\beta_{1}} =
\frac{d+2-\eta}{d-2+\eta_{\parallel}}, \nonumber\\
&& \delta_{11} = \frac{\Delta_{1}}{\beta_{1}}=
\frac{d-\eta_{\parallel}}{d-2+\eta_{\parallel}}\;.\nonumber
\end{eqnarray}

Each of these critical exponents characterizes certain properties
of the cubic anisotropic system near the surface. The values
$\nu$, $\eta$, and $\Delta=\nu(d+2-\eta)/2$ are the standard bulk
exponents.

\newpage
\begin{table}[htb]
\caption{Surface critical exponents of the ordinary transition for
$d=3$ up to two-loop order at the cubic fixed point (of order
$p=3$): $u^{*}=1.348, v^{*}=0.074$, at $n=3$.} \label{tab1}
\begin{center}
\begin{tabular}{rrrrrrrrrrr}
\hline
$ exp $~&~$\frac{O_{1}}{O_{2}}$~&~$\frac{O_{1i}}{O_{2i}}$~&~$[0/0]$~&~
$[1/0]$~&~$[0/1]$~&~$[2/0]$~&~$[0/2]$~&~$[11/1]$~&~$[1/11]$~&~$
f $ \\
\hline $\eta_{\parallel}$ & 2.27 & 1.67 & 2.00 &
1.681 & 1.725 & 1.541 & 1.594 & 1.429 & 1.428 & 1.429 \\

$\eta_{\perp}$ & 2.74 & 1.91 & 1.00 & 0.841 & 0.863 & 0.783 & 0.805 &
0.749 & 0.749 & 0.749 \\

$\Delta_{1}$ & 2.32 & 3.69 & 0.25 & 0.409 & 0.440 & 0.478 & 0.504 & 0.530
 & 0.530 & 0.530 \\

$\beta_{1}$ & -4.39 & -2.58 & 0.75 & 0.909 & 0.940 & 0.873 & 0.858 &
0.880 & 0.880 & 0.880 \\

$\gamma_{11}$ & 0.00 & 0.00 & -0.50 & -0.50 & -0.50 & -0.433 & -0.428
& --- & --- & -0.400 \\

$\gamma_{1}$ & 3.38 & 17.70 & 0.50 & 0.739 & 0.814 & 0.810 &
0.838 & 0.839 & 0.837 & 0.838 \\

$\delta_{1}$ & 1.99 & 2.53 & 1.67 & 1.844 & 1.865 & 1.933 & 1.957 &
2.023 & 2.023 & 2.023 \\

$\delta_{11}$ & 1.83 & 2.47 & 0.33 & 0.475 & 0.498 & 0.552 & 0.582 &
0.647 & 0.647 & 0.647 \\

\end{tabular}
\end{center}
\end{table}

\begin{table}[htb]
\caption{Surface critical exponents of the ordinary transition for
$d=3$ up to two-loop order at the cubic fixed point (of order
$p=6$) $u^{*}=1.321(18), v^{*}=0.096(20)$, at $n=3$.} \label{tab2}
\begin{center}
\begin{tabular}{rrrrrrrrrrr}
\hline
$ exp $~&~$\frac{O_{1}}{O_{2}}$~&~$\frac{O_{1i}}{O_{2i}}$~&~$[0/0]$~&~
$[1/0]$~&~$[0/1]$~&~$[2/0]$~&~$[0/2]$~&~$[11/1]$~&~$[1/11]$~&~$
f $ \\
\hline $\eta_{\parallel}$ & 2.29 & 1.68 & 2.00 &
1.684 & 1.727 & 1.545 & 1.597 & 1.436 & 1.435 & 1.436 \\

$\eta_{\perp}$ & 2.76 & 1.92 & 1.00 & 0.842 & 0.863 & 0.785 & 0.806 &
0.752 & 0.752 & 0.752 \\

$\Delta_{1}$ & 2.34 & 3.72 & 0.25 & 0.408 & 0.438 & 0.476 & 0.501 & 0.526
 & 0.526 & 0.526 \\

$\beta_{1}$ & -4.42 & -2.60 & 0.75 & 0.908 & 0.938 & 0.872 & 0.858 &
0.879 & 0.879 & 0.879 \\

$\gamma_{11}$ & 0.00 & 0.00 & -0.50 & -0.50 & -0.50 & -0.433 & -0.428
& --- & --- & -0.424 \\

$\gamma_{1}$ & 3.41 & 17.87 & 0.50 & 0.737 & 0.811 & 0.807 &
0.834 & 0.835 & 0.836 & 0.836 \\

$\delta_{1}$ & 2.01 & 2.55 & 1.67 & 1.842 & 1.863 & 1.930 & 1.953 &
2.018 & 2.017 & 2.018 \\

$\delta_{11}$ & 1.84 & 2.48 & 0.33 & 0.474 & 0.497 & 0.550 & 0.579 &
0.642 & 0.642 & 0.642 \\

\end{tabular}
\end{center}
\end{table}

\begin{table}[htb]
\caption{Surface critical exponents of the ordinary transition for
$d=3$ up to two-loop order at the cubic fixed point (of order
$p=2$) $ u^{*}=1.064, v^{*}=0.520$, at $n=4$.} \label{tab3}
\begin{center}
\begin{tabular}{rrrrrrrrrrr}
\hline
$ exp $~&~$\frac{O_{1}}{O_{2}}$~&~$\frac{O_{1i}}{O_{2i}}$~&~$[0/0]$~&~
$[1/0]$~&~$[0/1]$~&~$[2/0]$~&~$[0/2]$~&~$[11/1]$~&~$[1/11]$~&~$
f $ \\
\hline
$\eta_{\parallel}$ & 2.12 & 1.54 & 2.00 & 1.647 & 1.700 & 1.481 &
1.550 & 1.319 & 1.314 & 1.317 \\

$\eta_{\perp}$ & 2.51 & 1.74 & 1.00 & 0.824 & 0.850 & 0.753 & 0.783 &
0.705 & 0.705 & 0.705 \\

$\Delta_{1}$ & 2.07 & 3.27 & 0.25 & 0.426 & 0.464 & 0.511 & 0.549 & 0.588
 & 0.589 & 0.589 \\

$\beta_{1}$ & -4.95 & -2.64 & 0.75 & 0.926 & 0.964 & 0.891 & 0.873 &
0.898 & 0.900 & 0.899 \\

$\gamma_{11}$ & 0.0 & 0.0 & -0.50 & -0.50 & -0.50 & -0.409 & -0.400
& --- & --- & 0.311 \\

$\gamma_{1}$ & 2.91 & 12.61 & 0.50 & 0.765 & 0.860 & 0.856 &
0.900 & 0.899 & 0.902 & 0.901 \\

$\delta_{1}$ & 1.83 & 2.34 & 1.67 & 1.863 & 1.889 & 1.969 & 2.003 &
2.105 & 2.103 & 2.104 \\

$\delta_{11}$ & 1.70 & 2.31 & 0.33 & 0.490 & 0.519 & 0.583 & 0.623 &
0.727 & 0.724 & 0.726 \\

\end{tabular}
\end{center}
\end{table}

\begin{table}[htb]
\caption{Surface critical exponents of the ordinary transition for
$d=3$ up to two-loop order at the cubic fixed point (of order
$p=2$) $u^{*}=0.525, v^{*}=1.146$, at $n=8$.} \label{tab4}
\begin{center}
\begin{tabular}{rrrrrrrrrrr}
\hline
$ exp $~&~$\frac{O_{1}}{O_{2}}$~&~$\frac{O_{1i}}{O_{2i}}$~&~$[0/0]$~&~
$[1/0]$~&~$[0/1]$~&~$[2/0]$~&~$[0/2]$~&~$[11/1]$~&~$[1/11]$~&~$
f $ \\
\hline
$\eta_{\parallel}$ & 2.13 & 1.55 & 2.00 & 1.645 & 1.699 & 1.479 &
1.548 & 1.307 & 1.297 & 1.302 \\

$\eta_{\perp}$ & 2.51 & 1.74 & 1.00 & 0.823 & 0.849 & 0.752 & 0.781 &
0.702 & 0.701 & 0.702 \\

$\Delta_{1}$ & 2.05 & 3.22 & 0.25 & 0.427 & 0.466 & 0.514 & 0.553 & 0.592
 & 0.593 & 0.593 \\

$\beta_{1}$ & -5.45 & -2.77 & 0.75 & 0.927 & 0.966 & 0.895 & 0.878 &
0.903 & 0.905 & 0.904 \\

$\gamma_{11}$ & 0.0 & 0.0 & -0.50 & -0.50 & -0.50 & -0.393 & -0.380
& --- & --- & -0.315 \\

$\gamma_{1}$ & 2.84 & 11.57 & 0.50 & 0.766 & 0.863 & 0.860 &
0.907 & 0.902 & 0.909 & 0.906 \\

$\delta_{1}$ & 1.84 & 2.34 & 1.67 & 1.864 & 1.890 & 1.971 & 2.005 &
2.113 & 2.139 & 2.126 \\

$\delta_{11}$ & 1.70 & 2.33 & 0.33 & 0.491 & 0.521 & 0.584 & 0.624 &
0.739 & 0.732 & 0.736 \\

\end{tabular}
\end{center}
\end{table}

\begin{table}[htb]
\caption{Surface critical exponents of the special transition for
$d=3$ up to two-loop order at the cubic fixed point (of order
$p=2$) $u^{*}=0.201, v^{*}=1.508$, for $n\to \infty$.}
\label{tab5}
\begin{center}
\begin{tabular}{rrrrrrrrrrr}
\hline
$ exp $~&~$\frac{O_{1}}{O_{2}}$~&~$\frac{O_{1i}}{O_{2i}}$~&~$[0/0]$~&~
$[1/0]$~&~$[0/1]$~&~$[2/0]$~&~$[0/2]$~&~$[11/1]$~&~$[1/11]$~&~$
f $ \\
\hline
$\eta_{\parallel}$ & 2.14 & 1.56 & 2.00 & 1.648 & 1.701 & 1.484 &
1.552 & 1.312 & 1.300 & 1.306 \\

$\eta_{\perp}$ & 2.53 & 1.75 & 1.00 & 0.824 & 0.850 & 0.755 & 0.783 &
0.706 & 0.704 & 0.705 \\

$\Delta_{1}$ & 2.07 & 3.26 & 0.25 & 0.426 & 0.463 & 0.511 & 0.549 & 0.585
 & 0.587 & 0.586 \\

$\beta_{1}$ & -5.34 & -2.75 & 0.75 & 0.926 & 0.963 & 0.893 & 0.876 &
0.901 & 0.904 & 0.903 \\

$\gamma_{11}$ & 0.0 & 0.0 & -0.50 & -0.50 & -0.50 & -0.387 & -0.373
& --- & --- & -0.323 \\

$\gamma_{1}$ & 2.88 & 11.93 & 0.50 & 0.764 & 0.858 & 0.856 &
0.901 & 0.895 & 0.902 & 0.899 \\

$\delta_{1}$ & 1.85 & 2.36 & 1.67 & 1.862 & 1.888 & 1.968 & 2.001 &
2.108 & 2.105 & 2.107 \\

$\delta_{11}$ & 1.71 & 2.34 & 0.33 & 0.490 & 0.519 & 0.581 & 0.621 &
0.736 & 0.728 & 0.732 \\

\end{tabular}
\end{center}
\end{table}


\begin{thebibliography}{10}
\bibitem{mit3}
This is an intuitive picture conveyed from the theory of bulk
regular systems. Much less is known about the large-order behavior of
perturbative expansions pertaining to infinite random systems (see Refs.
  \cite{B87,Mc94,AMR00}), especially at large space dimensionalities.
At the present time, there are no explicit results on large orders
for the surface quantities, even in the absence of disorder.
\bibitem{mitSh88} We applied the formulas of $\beta$ functions, presented
in \cite{Sh88} for the case of cubic anisotropic model with $m=1$ and
$n=3$.
\bibitem{mit2} The values of $\nu$ and $\eta$ at $n=4,8$ and $n\to\infty$
are calculated from formulas, presented in \cite{Sh88} for the case of cubic
anisotropic model.
\bibitem{mit5} In order to evaluate the difference between surface critical
exponents of semi-infinite model with cubic anisotropy and surface critical
exponents of standard semi-infinite $n$-component model
\cite{DD81,DD83,DSh94,DSh98} we performed additional calculation for surface
critical exponents on the basis of formulas from \cite{DSh94,DSh98} in the
case of 3D semi-infinite model with $n=3$.
\bibitem{mit4} The investigation of the surface
critical behavior of random semi-infinite systems at the special and
ordinary surface transition was performed early in serial works
\cite{UShH00,UH02}. The obtained results indicate that the surface critical
exponents of random semi-infinite systems belong to the universality class
of random fixed point.
\end{thebibliography}
\end{document}